\documentclass{article}
\usepackage{spconf}
\usepackage{amsmath}
\usepackage{graphicx, graphics, theorem, times, amsfonts, amsmath, amssymb, cite}
\usepackage{tikz}
\usetikzlibrary{shapes,arrows}
\usepackage{pgfplots}
\usepackage{color}
\usepackage{hyperref}
\usepackage{multirow}
\usepackage{rotating}
\usepackage{cleveref, subcaption}
\usepackage{bm}
\input{mysymbol.sty}
\usepackage[utf8]{inputenc}
\usepackage{enumitem}

\newcommand{\EE}{{\mathbb E}}

\newcommand{\QED}{\hfill\ensuremath{\blacksquare}}

\ninept

\newtheorem{problem}{\bf Problem}

\def \bbCo {\bbC_{{\scriptscriptstyle \ccalO}}}

\def \bbCoh {\bbC_{{\scriptscriptstyle \ccalO\ccalH}}}

\def \hbCo {\hbC_{{\scriptscriptstyle \ccalO}}}
\def \hbCh {\hbC_{{\scriptscriptstyle \ccalH}}}
\def \hbCoh {\hbC_{{\scriptscriptstyle \ccalO\ccalH}}}

\def \bbXo {\bbX_{{\scriptscriptstyle \ccalO}}}

\def \bbSo {\bbS_{{\scriptscriptstyle \ccalO}}}
\def \bbSh {\bbS_{{\scriptscriptstyle \ccalH}}}
\def \bbSoh {\bbS_{{\scriptscriptstyle \ccalO\ccalH}}}

\def \hbSo {\hbS_{{\scriptscriptstyle \ccalO}}}

\newcommand{\norm}[1]{\left\lVert#1\right\rVert}

\title{Joint inference of multiple graphs with hidden variables from stationary graph signals}
\name{Samuel Rey$^\star$, Andrei Buciulea$^\star$, Madeline Navarro$^\dagger$, Santiago Segarra$^\dagger$, and Antonio G. Marques$^\star$
\thanks{Work supported by the Spanish Fed. Grants FPU17-04520, EST21/00420, and SPGraph PID2019-105032GB-I00, and by the US NSF under award CCF-2008555.}}
\address{ $^\star$Dept. of Signal Theory and Communications, King Juan Carlos University, Madrid, Spain \\ $^\dagger$Dept. of Electrical and Computer Engineering, Rice University, Houston, USA}

\begin{document}
%\ninept
%
\maketitle
\begin{abstract}
Learning graphs from sets of nodal observations represents a prominent problem formally known as graph topology inference.
However, current approaches are limited by typically focusing on inferring single networks, and they assume that observations from all nodes are available.
First, many contemporary setups involve multiple related networks, and second, it is often the case that only a subset of nodes is observed while the rest remain hidden.
Motivated by these facts, we introduce a joint graph topology inference method that models the influence of the hidden variables.
Under the assumptions that the observed signals are stationary on the sought graphs and the graphs are closely related, the joint estimation of multiple networks allows us to exploit such relationships to improve the quality of the learned graphs.
Moreover, we confront the challenging problem of modeling the influence of the hidden nodes to minimize their detrimental effect.
To obtain an amenable approach, we take advantage of the particular structure of the setup at hand and leverage the similarity between the different graphs, which affects both the observed and the hidden nodes.
To test the proposed method, numerical simulations over synthetic and real-world graphs are provided.

% Present our solution
% Talk about experiments
\end{abstract}
\section{Introduction}\label{sec:introudction}
% Importance of graphs
Graphs have been successfully exploited to capture the irregular (non-Euclidean) structure commonly inherent to contemporary data for several years now.
Increasingly often, several disciplines such as statistics, machine learning, or signal processing (SP), among others, rely on graphs to capture the underlying irregular domain for solving a range of applications on, e.g., communications, genetics, and brain networks~\cite{kolaczyk2009book,sporns2012book,EmergingFieldGSP,ortega_2018_graph}.
% Introducing topology inference problem
However, despite the growing popularity of graph-related methods, in many situations the graph is unknown and we only have access to a set of nodal observations.
Then, under the core assumption that the properties of the nodal observations and the topology of the sought graph are closely related, it is possible to learn the network based on the observed signals.
This constitutes a prominent problem commonly known as \emph{graph topology inference}~\cite{mateos2019connecting,sardellitti2019graph}.
Notable approaches include correlation networks~\cite[Ch. 7.3.1]{kolaczyk2009book}, partial correlations and (Gaussian) Markov random fields~\cite{meinshausen06,GLasso2008,kolaczyk2009book}, sparse structural equation models~\cite{BainganaInfoNetworks}, graph-SP-based approaches~\cite{MeiGraphStructure,egilmez2017graph,segarra2017network,segarra2018network}, as well as their non-linear generalizations\cite{Karanikolas_icassp16}.
% Removed cites: GMRF --> Lake10discoveringstructure, SSEM --> BazerqueGeneNetworks

% Joint topology inference and hidden variables
A common denominator of the previous works is that: (i) they focus on identifying a single network; and (ii) they assume that observations (measurements) from all the nodes are available.
It is relevant to address the first item because many contemporary setups involve \emph{multiple related networks}, each of them with a subset of available signals.
This is the case, for example, in multi-hop communication networks in dynamic environments, in social networks where the same set of users may present different types of interactions, or in brain analytics where observations from different patients are available and the goal is to estimate their brain functional networks. 
When there exist several closely related networks, we can boost the performance of network topology inference by approaching the problem in a joint fashion that allows us to capture the relationship between the different graphs~\cite{murase2014multilayer,danaher2014joint,navarro2020joint,arroyo2021inference,wang_2020_high}.
Regarding the second point, assuming that observations from the whole graph are available may not always be realistic.
In fact, in many relevant settings the \emph{observed} signals may correspond only to a subset of the nodes from the original graph while the rest of them remain \emph{hidden}. 
If these hidden nodes are not properly accounted for, they can drastically hinder the performance of the network topology inference methods.
Therefore, the presence of hidden variables entails a challenge for most of the existing algorithms, and they require important adjustments.
Some works that are starting to deal with this relevant topic include Gaussian graphical model selection \cite{chandrasekaran2012latent,chang2019graphical}, inference of linear Bayesian networks \cite{anandkumar2013learning}, nonlinear regression \cite{mei2018silvar}, and our previous work based on graph stationarity \cite{buciulea2019network,buciulea2021learning}, to name a few.

% Our contribution
Based on the previous discussion, the contribution of this paper is to propose a topology inference method that simultaneously performs joint estimation of \emph{multiple} graphs and accounts for the presence of hidden variables.
To that end, we rely on results from graph SP (GSP), an area in SP that emerged as a way to generalize tools originally conceived to process signals with regular supports and extend them to signals defined in heterogeneous domains represented by a graph~\cite{EmergingFieldGSP,SandryMouraSPG_TSP13,ortega_2018_graph}.
We assume that the observed signals constitute a random process that is \emph{stationary} on the given graph.
As done in our previous work~\cite{buciulea2021learning}, to formalize the relationship between the unknown network and the nodal observations, the graph stationarity assumption must be modified to model the influence of the hidden nodes.
Then, to fully benefit from the joint inference formulation, a critical aspect is to capture the similarity among graphs not only accounting for the observed nodes but also for the hidden ones.
This is achieved by carefully exploiting the structure inherent to the presence of latent variables with a regularization inspired by group Lasso~\cite{simon2013sparse}.
The proposed method is evaluated using synthetic and real-world graphs and compared with other related approaches.

% Outline
The remainder of the paper is organized as follows.
Section~\ref{preliminaries} introduces some GSP concepts leveraged during the paper, Section~\ref{graph_filter_id} formally introduces the problem at hand and presents the proposed method, and Section~\ref{results} offers a numerical evaluation.
Finally, Section~\ref{conclusions} provides some closing remarks.

\section{Preliminaries} \label{preliminaries}
We now introduce basic GSP concepts that will help in explaining the relation between the unknown graph and the observed signals while modeling the influence of the hidden variables.

\noindent\textbf{Graph signal processing.}
Let $\ccalG=(\ccalN,\ccalE)$ be an undirected graph where $\ccalN$ is the set of nodes with cardinality $|\ccalN|=N$, and $\ccalE$ is the set of edges such that $(i,j)\in\ccalE$ only if nodes $i$ and $j$ are connected.
For a given $\ccalG$, the adjacency matrix $\bbA \in \reals ^{N\times N}$ is a (sparse) matrix with non-zero elements $A_{ij}$ if and only if $(i,j)\in\ccalE$.
Related to $\bbA$ is the GSO, a square matrix that captures the topology of the graph.
The GSO is defined as a matrix $\bbS\in\reals^{N \times N}$ whose entry $S_{ij}$ can be non-zero only if $i=j$ or $(i,j)\in\ccalE$. 
Typical choices for $\bbS$ are the adjacency matrix $\bbA$ and the graph Laplacian $\bbL$, which is defined as  $\bbL:=\diag(\bbA\bbone)-\bbA$ \cite{EmergingFieldGSP,djuric2018cooperative}.
We assume that $\bbS$ is diagonalizable so it can be represented as $\bbS=\bbV\bbLambda\bbV^T$, where $\bbV$ is an $N \times N$ orthogonal matrix collecting the eigenvectors and $\bbLambda$ is a diagonal matrix collecting the eigenvalues of $\bbS$.
Then, signals defined on the nodes of $\ccalG$ are called \textit{graph signals}, which are functions $f : \mathcal{N} \mapsto \mathbb{R}$, equivalently represented as vectors $\mathbf{x}=[x_1,...,x_N]^T \in  \mathbb{R}^N$, where $x_i$ denotes the signal value at node $i$. 
A key assumption of GSP is that since graph signals are defined on top of the graph, their statistical properties are closely related to the topology of $\ccalG$.
A fundamental tool when dealing with graph signals are \textit{graph filters}, linear graph-signal operators that account for the topology of the graph and can be defined as polynomials of the GSO $\bbS$, i.e., $\bbH \!=\!\!\sum_{l=0}^{L-1}\! h_l\bbS^l$, where $\bbh=[h_0,...,h_L]^T$ is the vector collecting the filter coefficients.
When applied to an input graph signal $\bbx$, the output of the graph filter is $\bby=\bbH\bbx =\sum_{l=0}^{L-1} h_l (\bbS^l\bbx)$, where $\bbS^l\bbx$   can be viewed as the diffusion of $\bbx$ across an $l$-hop neighborhood, and $h_l$ are the coefficients of the linear combination~\cite{segarra2017optimal}.

\noindent\textbf{Graph stationarity.}
A \textit{random} graph signal $\bbx$ with zero mean and covariance $\bbC \!= \!\EE[\bbx\bbx^T]$ is said to be stationary in the underlying graph $\ccalG$ if its covariance matrix $\bbC$ is diagonalized by $\bbV$, the eigenvectors of the GSO $\bbS$~\cite{djuric2018cooperative}.
Equivalently\footnote{A small technical condition must hold for these two statements to be equivalent; see~\cite{marques2017stationary}.}, a \textit{random} graph process is defined to be stationary in $\ccalG$ if it can be represented as the output of a graph filter $\bbH$, which is a matrix polynomial in $\bbS$, to a white input.
Specifically, under the stationarity assumption the process $\bbx$ can be written as $\bbx=\bbH\bbw$, where $\bbw$ is a random vector of white noise.
When particularized to discrete time-varying signals, the two aforementioned definitions boil down to the classical definition of stationarity in time~\cite{marques2017stationary}.
Also note that graph stationarity implies that the covariance of $\bbx$ and the GSO commute, so we have that
\begin{equation}\label{E:commutativity}
    \bbC\bbS=\bbS\bbC.
\end{equation}

\section{Joint inference in the presence of hidden variables}\label{graph_filter_id}
% Introduce the notation
To formally introduce the problem of joint graph topology inference in the presence of hidden variables, let us assume that $K$ undirected graphs $\{\ccalG^{(k)}\}_{k=1}^K$ are defined over the same set of nodes $\ccalN$, and denote as $\bbX^{(k)}=[\bbx_1^{(k)},...,\bbx_{M_k}^{(k)}]\in\reals^{N \times M_k}$ the collection of (zero-mean) $M_k$ signals defined on top of each unknown graph $\ccalG^{(k)}$.
Furthermore, consider that for each graph only a subset of nodes $\ccalO \subset \ccalN$ with cardinality $O<N$ is observed, while the remaining $H$ nodes in the subset $\ccalH=\ccalN \setminus \ccalO$ stay hidden.
Without loss of generality, let the signals associated with the observed nodes be collected in the first $O$ rows of $\bbX^{(k)}$ and denote them as $\bbXo^{(k)} \in \reals^{O \times M_k}$.
Then, it can be seen that, for each graph, the unknown GSO $\bbS^{(k)}$ and the sample covariance  $\hbC^{(k)}=\frac{1}{M_k}\bbX^{(k)}(\bbX^{(k)})^T$ are symmetric matrices with the following block structure
\begin{equation}\label{E:block_SC}
    \bbS^{(k)} =
    \begin{bmatrix}
    \bbSo^{(k)} & \bbSoh^{(k)}  \\
    (\bbSoh^{(k)})^T & \bbSh^{(k)}
    \end{bmatrix}\!, \;
    \hbC^{(k)} =
    \begin{bmatrix}
    \hbCo^{(k)} & \hbCoh^{(k)}  \\
    (\hbCoh^{(k)})^T & \hbCh^{(k)}
    \end{bmatrix}. \;
\end{equation}
The $O \times O$ matrices $\bbSo^{(k)}$ denote the block of the GSOs capturing the connections between the observed nodes while the rest of the submatrices involve edges connected to hidden nodes.
Similarly, $\hbCo^{(k)}$ denotes the sample covariance of the observed signals $\bbXo^{(k)}$.

% Problem formulation
With these considerations in place, the problem of joint topology inference in the presence of hidden variables is introduced next.
\begin{problem}\label{general_problem}
    Given the $O \times M_k$ matrices $\{\bbXo^{(k)}\}_{k=1}^K$ collecting the signal values at the observed nodes for each graph $\ccalG^{(k)}$, find the sparsest matrices $\{\bbSo^{(k)}\}_{k=1}^K$ encoding the structure of the $K$ graphs under the assumptions that: \\
    (AS1) The number of hidden nodes is much smaller than the number of observed nodes, i.e., $ H \ll O$; \\
    (AS2) The signals $\bbX^{(k)}$ are realizations of a random process that is stationary in $\bbS^{(k)}$; and \\
    (AS3) The distance between the $K$ graphs is small according to a particular metric $d(\bbS^{(k)},\bbS^{(k')})$. 
\end{problem}
%

% Explain the problem
Accounting for the hidden variables implies modeling their influence over the observed nodes without any additional observation, thus rendering the inference problem a challenge.
To ensure the tractability of the problem, (AS1) ensures that most of the nodes are observed while (AS2) establishes a relation between the graph signals and the whole unknown graph, including the hidden nodes.
Then, (AS3) guarantees that the $K$ graphs are related so we can benefit from inferring them in a joint setting.

% Introduce next section
In the following section, we exploit the aforementioned assumptions and the block structure resulting from the presence of hidden variables to approach Problem~\ref{general_problem} by solving a convex optimization problem. 
 
\subsection{Modeling hidden variables in the joint inference problem}
% Influence of hidden variables in stationarity
Fundamental to approach Problem~\ref{general_problem} is modeling the impact of the hidden nodes in the stationarity assumption (AS2), which implies that the matrices $\bbC^{(k)}$ and $\bbS^{(k)}$ are simultaneously diagonalized by the eigenvectors of the $k$th graph $\bbV^{(k)}$.
To that end, similar to the previous work in~\cite{buciulea2021learning}, we avoid the challenging task of obtaining the submatrix of eigenvectors corresponding to the observed nodes by leveraging the commutativity of the matrices $\bbC^{(k)}$ and $\bbS^{(k)}$, and the block structure introduced in \eqref{E:block_SC}.
More specifically, upon \emph{lifting} the $O \times O$ matrices $\bbP^{(k)}:=\bbCoh^{(k)}(\bbSoh^{(k)})^T$, if we focus on the upper left block at both sides of the equality in \eqref{E:commutativity} we have that the graph stationarity in the presence of hidden nodes is represented by
\begin{equation}\label{E:block_commutativity}
    \bbCo^{(k)}\bbSo^{(k)}+\bbP^{(k)}=\bbSo^{(k)}\bbCo^{(k)}+(\bbP^{(k)})^T.
\end{equation}
%
% low rank of P
Furthermore, the matrices $\bbP^{(k)}$ are the product of two matrices of sizes $O \times H$ and $H \times O$.
Then, due to (AS1) we have that $H \ll O$, and hence we know that the rank of $\bbP^{(k)}$ is upper bounded by $H$.
%Moreover, since $\bbSoh^{(k)}$ is a block of the GSO of a sparse graph $\ccalG^{(k)}$, it can be seen that the matrices $\bbP^{(k)}$ present a column-sparse structure.

% Non-convex formulation
With the previous considerations in place, we approach the sparse joint topology inference problem in the presence of hidden nodes by means of the following non-convex optimization problem
\begin{alignat}{2}\label{E:eqn_zero_norm}
    \!\!&\!\min_{\{\bbSo^{(k)},\bbP^{(k)}\}_{k=1}^K} \
    &&\sum_k \alpha_k \|\bbSo^{(k)}\|_0 + \sum_{k<k'}\beta_{k,k'}d_S(\bbSo^{(k)},\bbSo^{(k')})                                   \\ 
    \!\!&\! && + \sum_{k<k'}\eta_{k,k'}d_P(\bbP^{(k)},\bbP^{(k')}) \nonumber \\
    \!\!&\!\mathrm{\;\;s. \;t. } && \;\;\;\;\rank(\bbP^{(k)})\leq H, \;\; \nonumber
    \\
    \!\!&\! && 
    \;\;\;\; \|\hbCo^{(k)}\bbSo^{(k)}\!+\!\bbP^{(k)} \!-\!\bbSo^{(k)}\hbCo^{(k)}\!-\!(\bbP^{(k)})^{T}\|_F^2 \leq  \epsilon, \;\; \nonumber
    \\
    \!\!&\! && 
    \;\;\;\; \bbSo^{(k)} \in \ccalS. \;\; \nonumber
\end{alignat}
The first and second constraints capture assumptions (AS1) and (AS2), with $\epsilon$ being a small positive number capturing the fidelity of the sample covariance. The set $\ccalS$ encodes the properties of the desired GSOs.
In this paper we will focus on the case where the GSO is given by the adjacency matrix of the underlying undirected graph with non-negative weights and no self-loops.
Thus, from now onwards we set the feasibility set $\ccalS$ to be given by
\begin{equation}\label{E:feasibility_sets}
    \ccalS = \ccalS_{\mathrm{A}} \!:= \! \{ \bbS \, | \, S_{ij} \geq 0, \;\,  \bbS = \bbS^T,\;\,  S_{ii} = 0, \;\, \textstyle\sum_j S_{j1} \! = \! 1 \}, \nonumber
\end{equation}
where the last condition fixes the scale of the admissible graphs by setting the weighted degree of the first node to $1$, which rules out the trivial solution $\bbS\!=\!\bbzero$.
Other GSOs such as the normalized Laplacian can be accommodated via minor adaptations to $\ccalS$; see~\cite{segarra2017network}.

% Effect of the joint inference in the hidden variables
Similar to standard joint inference approaches~\cite{navarro2020joint}, the objective function of \eqref{E:eqn_zero_norm} captures the similarity of the $K$ graphs with the function $d_S(\cdot,\cdot)$.
Nevertheless, when accounting for the presence of hidden variables, assumption (AS3) is also reflected in the unobserved blocks of the GSOs.
This important observation, captured by the function $d_P(\cdot,\cdot)$, allows us to incorporate additional structure reducing the degrees of freedom and rendering the problem more manageable.
More specifically, note that the matrix $\bbP^{(k)}$ is given by the product of $\bbCoh^{(k)}$ and $(\bbSoh^{(k)})^T$ with the latter being a submatrix of a sparse  GSO, so it can be seen that the matrices $\bbP^{(k)}$ present a column-sparse structure.
Furthermore, since the $K$ graphs are similar, the submatrices $\bbSoh^{(k)}$ are also similar, which implies that the matrices $\bbP^{(k)}$ present a similar column-sparsity pattern.
In other words, the columns with non-zero entries are likely to be placed in the same positions for the different matrices $\bbP^{(k)}$.
By designing a distance function $d_P(\cdot,\cdot)$ that exploits this additional structure we improve the estimation of the matrices $\bbP^{(k)}$, resulting in a better estimation of the matrices $\bbSo^{(k)}$. 

% Intro next section
The non-convexity of \eqref{E:eqn_zero_norm}, which arises from the presence of the rank constraint and the $\ell_0$ norm, renders the optimization problem computationally hard to solve, leading us to implement some convex relaxations that are detailed next.

\subsection{Convex relaxations for the joint topology inference}
% Convex relaxation --> group lasso and reweighted
The rank constraints are commonly avoided by augmenting the objective function with a nuclear norm penalty, which promotes low-rank solutions by seeking matrices with sparse singular values. 
However, this penalty does not preserve the characteristic column sparsity of the matrices $\bbP^{(k)}$.
To circumvent this issue, in contrast to~\cite{buciulea2021learning}, we employ the group Lasso regularization~\cite{simon2013sparse} and rely on the fact that, in this particular setting, we can promote low rankness by reducing the number of non-zero columns while still achieving a reliable estimate.
Then, we replace the $\ell_0$ norm by a reweighted $\ell_1$ minimization~\cite{candes2008enhancing}, an iterative algorithm rooted on a logarithmic penalty that usually outperforms the widely used $\ell_1$ norm.

% Our algorithm
%% - Comment on convergence of the algorithm?
By leveraging the aforementioned relaxations we address the joint topology inference problem in the presence of hidden variables by solving an iterative method.
Under this approach, for each iteration, we solve the following convex problem
\begin{alignat}{2}\label{E:eqn_convex}
    \!\!&\!  \min_{\scriptscriptstyle\{\bbSo^{(k)},\bbP^{(k)}\}_{k=1}^K} &&
    \;\sum_k \!\alpha_k \mathrm{vec}(\bbW^{(k)})^T\!\mathrm{vec}(\bbSo^{(k)})
    \\
    \!\!&\!  && \!\!\!+\! \sum_{k<k'}\!\beta_{k,k'}\|\bbSo^{(k)}-\bbSo^{(k')}\|_1 \! \nonumber
    \\ 
    \!\!&\!  && \!\!\!+\!  \sum_k\!\gamma_k\|\bbP^{(k)}\|_{2,1} +\! \sum_{k<k'}\eta_{k,k'} \norm{\begin{bmatrix}
        \bbP^{(k)} \\
        \bbP^{(k')}
    \end{bmatrix}}_{2,1} \; \nonumber
    \\
    \!\!&\!  && \!\!\!+\! \sum_k\!\mu_k \|\! \hbCo^{(k)}\bbSo^{(k)} \!+\! \bbP^{(k)} \!-\! \bbSo^{(k)}\hbCo^{(k)} \!-\! (\bbP^{(k)})^{T} \!\|_F^2 \nonumber
    \\
    \!\!&\!\mathrm{\;\;s. \;t. } && \bbSo^{(k)} \in \ccalS. \;\; \nonumber
\end{alignat}
%\red{AGM: Is $|\|\bbp_i^{(k)} \|_1 -\|\bbp_i^{(k')}\|_1|$ convex? I don't think so. It seems to me that one way to circumvent this is to drop that term, define the matrix $\bbP$ as the vertical concatenation of $\bbP^{(1)}...\bbP^{(K)}$ and include a new regularizer minimizing $\|\bbP\|_{2,1}$, so that we promote the same sparsity pattern across all the matrices  $\{\bbP^{(k)}\}_{k=1}^K$.}  
%
%where $\alpha_k,\beta_{k,k'},\gamma_k,\eta_{k,k'}$ and $\mu_k$ are positive numbers selected based on prior knowledge,
%where $\bbp_i^{(k)}$ denotes the $i$th column of the matrix $\bbP^{(k)}$, and weight matrices $\bbW^{(k)}$ whose computation is described next.
%To that end, 
To compute the weight matrices $\bbW^{(k)}$, let $t=1...T$ denote the iteration index (omitted in the expression above to alleviate the notation), and compute the $k$th weight matrix for the $t$th iteration as $W^{(k,t)}_{ij}=(S_{\scriptscriptstyle \ccalO_{ij}}^{(k,t-1)}+\delta)^{-1}$ with $S_{\scriptscriptstyle \ccalO_{ij}}^{(k,t-1)}$ being the solution obtained during the $t-1$th iteration and $\delta$ a small positive constant. 
Hence, for each iteration $t$ we first compute the weight matrices $\bbW^{(k,t)}$ and, then, employ those to estimate the matrices $\bbSo^{(k,t)}$ and $\bbP^{(k,t)}$.
Coming back to the formulation in \eqref{E:eqn_convex}, note that the distance $d_S(\cdot,\cdot)$ is set to the $\ell_1$ norm to promote similar edges on the $K$ graphs.
The norm $\|\cdot\|_{2,1}$ represents the group Lasso penalty by first computing the $\ell_2$ norm of the columns of the input matrix and then the $\ell_1$ norm of the resulting vector.
To capture the similar column-sparsity pattern of the matrices $\bbP^{(k)}$ resulting from the similarity of the $K$ graphs, we design the function $d_P(\cdot,\cdot)$ relying on the group Lasso penalty.
More specifically, we concatenate each pair of matrices $\bbP^{(k)}$ and $\bbP^{(k')}$ to create a tall matrix and then promote column sparsity on the tall matrix with the $\ell_{2,1}$ norm.
Note that a column of all zeros in the tall matrix implies that the same column in $\bbP^{(k)}$ and $\bbP^{(k')}$ will only contain zeros, thus promoting the desired structure.
Finally, it is worth noting that we moved the commutativity constraint to the objective function.
Due to the iterative nature of the proposed method, the estimation of the observed GSO during the first iteration might be far from the true GSO, and hence, a more restrictive constraint as the one employed in \eqref{E:eqn_zero_norm} might be misleading.
Augmenting the objective function with the commutativity penalty is more amenable to an iterative approach.

%%%%%%%%%   FIGURES   %%%%%%%%%%%%%%%
\begin{figure*}[!t]
	\centering
	\begin{subfigure}{0.32\textwidth}
		\centering
		    \includegraphics[width=1\textwidth]{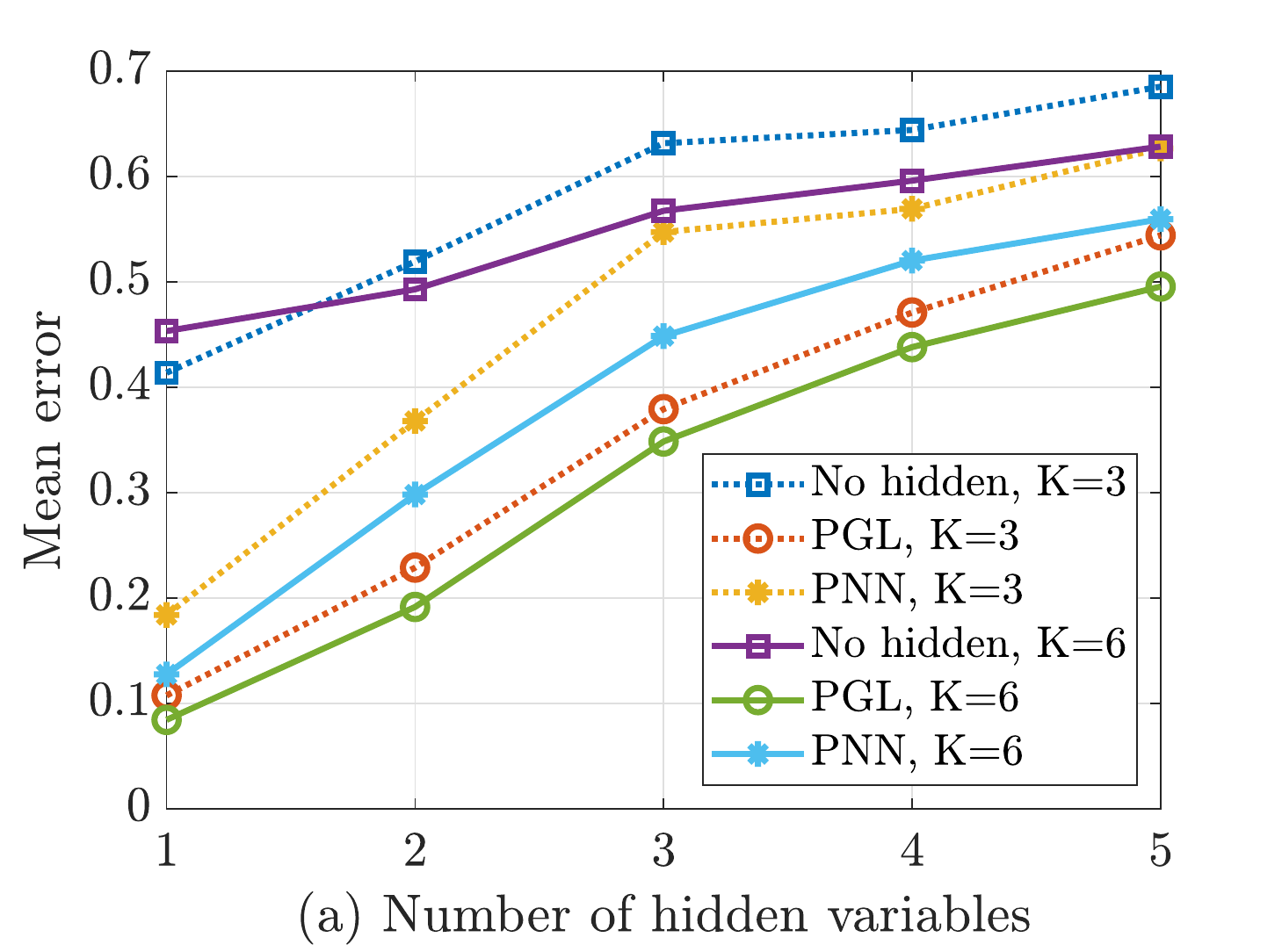}
	\end{subfigure}
	\begin{subfigure}{0.32\textwidth}
		\centering
			\includegraphics[width=1\textwidth]{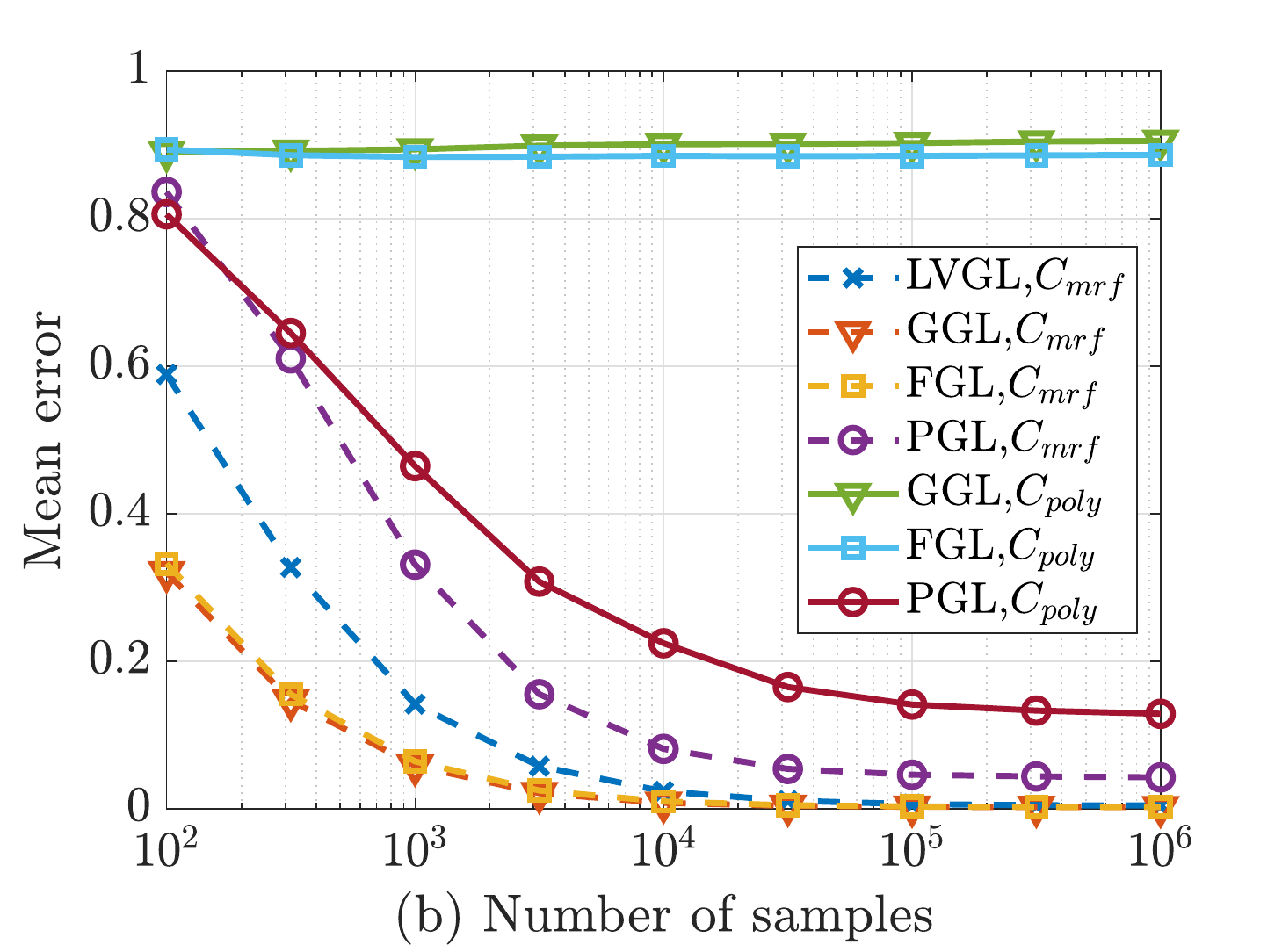}
	\end{subfigure}
	\begin{subfigure}{0.32\textwidth}
		\centering
			\includegraphics[width=1\textwidth]{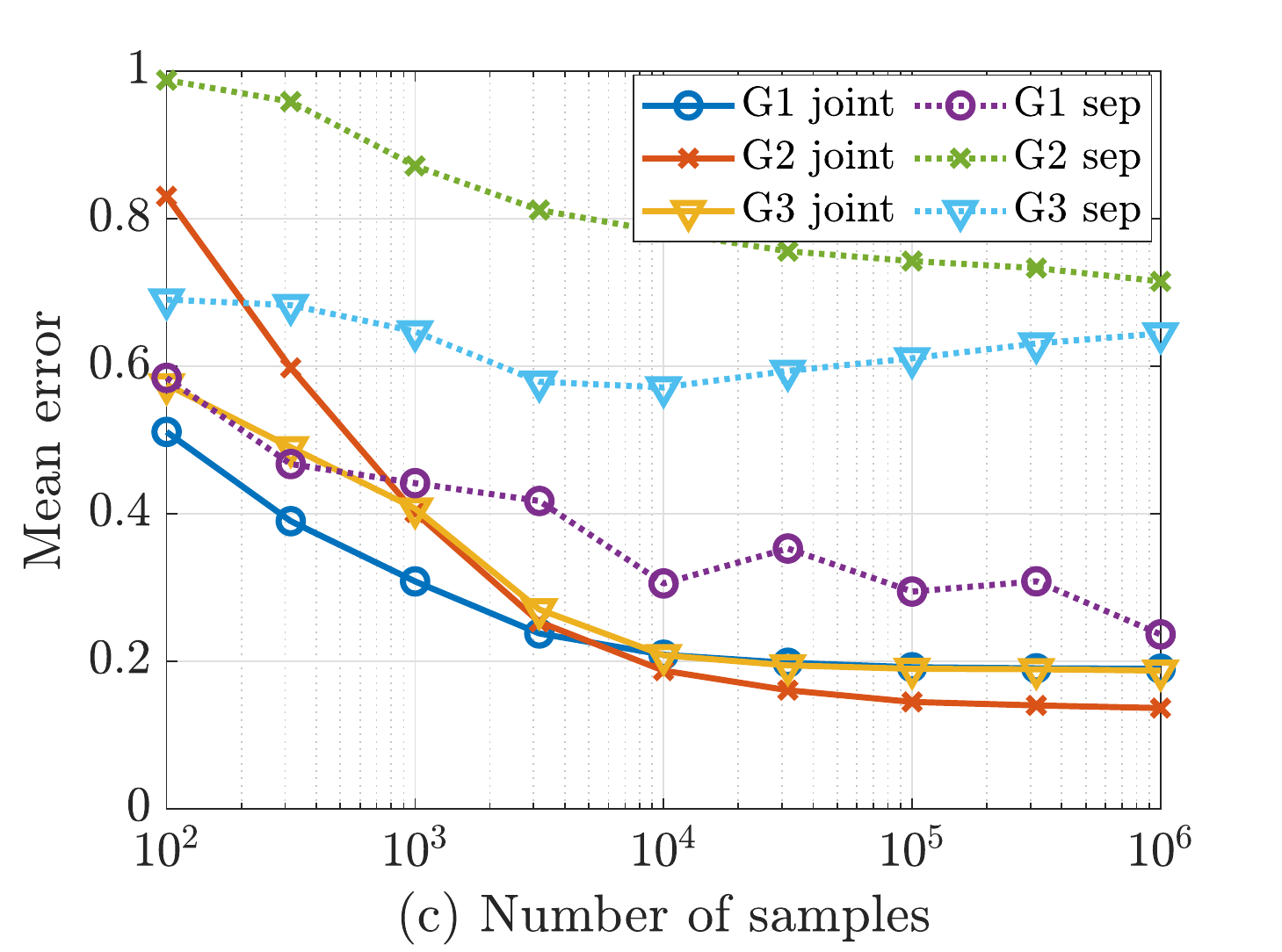}
	\end{subfigure}
	\vspace{-.25cm}
	\caption{Numerical validation of the proposed algorithm. 
	a)~Mean error of 100 realizations as the number of hidden variables increases for different models and values of $K$.
	b)~Mean error of the recovered graphs for several algorithms as the number of samples increases. 
	c)~Mean error of the recovered graphs for joint and separate approaches as the number of samples increases. 
	The first two experiments use Erd\H{o}s-Rényi graphs with $N=20$ and $p=0.2$ and the third one employs real-world graphs.} \label{fig:main_fig}
	\vspace{-.35cm}
\end{figure*}

\vspace{-0.5mm}
\section{Numerical experiments}\label{results}
We now introduce the simulations performed over synthetic and real-world graphs to evaluate the proposed method.
%The error of the recovered graphs is computed as $\sum_{k=1}^K\|\bbSo^{(k)}-\hbSo^{(k)}\|_F^2/K\|\bbSo^{(k)}\|_F^2$, and when graphs are randomly generated, they are sampled from an Erd\H{o}s-Rényi (ER) model with $N=20$ nodes and edge probability $p=0.2$.
When graphs are randomly generated, they are sampled from an  Erd\H{o}s-Rényi (ER) model with $N=20$ nodes and edge probability $p=0.2$.
The code for the following experiments is available on
GitHub\footnote{\url{https://github.com/reysam93/hidden_joint_inference/tree/ICASSP2022}} and the interested reader is referred there for specific implementation details.

\vspace{1mm}
\noindent\textbf{Test case 1.}
In the first experiment, we evaluate the influence of the hidden variables and its detrimental effect on the topology inference task
when the true covariance matrix is known.
The results are depicted in Figure~\ref{fig:main_fig}a, where we report the error of the recovered graphs, computed as $\sum_{k=1}^K\|\bbSo^{(k)}-\hbSo^{(k)}\|_F^2/K\|\bbSo^{(k)}\|_F^2$, for several models as the number of hidden variables increases on the x-axis.
The error is averaged over 64 realizations with ER graphs.
The considered models are: (i)~``PGL'', which stands for the method introduced in \eqref{E:eqn_convex}; (ii)~``PNN'', which denotes the reweighed algorithm proposed in~\cite{buciulea2019network} augmented with the joint penalty $d_S(\cdot,\cdot)$ to perform the joint inference; and (iii)~``No hidden'', which is a joint inference method unaware of the presence of hidden variables similar to the work in~\cite{navarro2020joint}.
In addition, for each model we let $K$ take the values in $\{3,6\}$.
Looking at the results, we can observe that ``PGL'' and ``PNN'', which take into account the presence of hidden variables, outperform the method ``No hidden'', showcasing the benefit of a robust formulation.
Also, the method proposed in \eqref{E:eqn_convex} outperforms ``PNN'', the other alternative accounting for hidden variables.
This reflects the advantage of employing the group Lasso regularization and incorporating the graph similarity through the careful design of the function $d_P(\cdot,\cdot)$. 
Lastly, it is worth noting that  the performance improves for higher values of $K$, achieving better results when more related graphs are available.
%This suggest that there is a benefit from following a joint inference approach, which is further studied in the third experiment.

\vspace{1mm}
\noindent\textbf{Test case 2.}
Next, we evaluate the influence of the number of observed signals and compare the performance of the proposed approach with other related alternatives.
In this experiment, only a single hidden node is considered.
To that end, in Figure~\ref{fig:main_fig}b we show the mean normalized error of the recovered graphs on the y-axis as the number of samples increases on the x-axis.
%The error is computed as $\sum_{k=1}^K\|\bbSo^{(k)}-\hbSo^{(k)}\|_F^2/K\|\bbSo^{(k)}\|_F^2$, and the mean is considered over 30 realizations of $K$ graphs with 10 realizations of random covariance matrices for each, resulting in a total of 300 realizations.
The error is computed as in the previous experiment and the mean is considered over 30 realizations of $K=3$ ER graphs with 10 realizations of random covariance matrices for each, resulting in a total of 300 realizations.
We compare the proposed model (``PGL'') with latent variable graphical Lasso (``LVGL'')\cite{chandrasekaran2012latent}, and with group and fusion graphical Lasso (``GGL'' and ``FGL''), both from \cite{danaher2014joint}.
For each model, signals are generated using two different types of covariance matrices: (i) $\bbC_{mrf}=(\sigma\bbI+\phi\bbS)^{-1}$ where $\phi$ is a positive random number and $\sigma$ is a positive number so that  $\bbC_{MRF}^{-1}$ is positive semi-definite; and $\bbC_{poly}=\bbH^2$ where the matrix $\bbH$ is a graph filter with random coefficients $\bbh$.
By looking at the Figure~\ref{fig:main_fig}b, it can be observed that, when $\bbC_{mrf}$ is employed, the graphical Lasso models slightly outperform the proposed approach. 
This is expected since they are tailored for this specific type of covariance matrices.
However, we can also see that the performance of the proposed algorithm is close to that of the alternatives, illustrating the benefits of considering both the joint optimization and the presence of hidden variables.
On the other hand, when we focus on the covariance matrices $\bbC_{poly}$, it is evident that the proposed method ``PGL'' clearly outperform the alternatives, demonstrating that the proposed method is based on more general assumptions.
%The fact that the proposed method achieve good results for both classes of covariance matrices illustrates the benefit of following a more general approach, since $\bbC_{mrf}$ represent a general instantiation of $\bbC_{poly}$. 
Note that the results for ``LVGL'' for the polynomial covariance are not included since the error was too high. 

\vspace{1mm}
\noindent\textbf{Test case 3.}
Finally, we test the proposed algorithm and the impact of performing the topology inference in a joint fashion using real-world graphs.
We employ three graphs defined on a common set of 32 nodes.
Nodes represent students from the University of Ljubljana and the different networks encode different types of interactions among the students\footnote{Original data can be found at \url{http://vladowiki.fmf.uni-lj.si/doku.php?id=pajek:data:pajek:students}}.
The error is computed as before and one hidden variable is considered.
The results, illustrated in Figure~\ref{fig:main_fig}c, show the error of the recovered graphs as the number of samples increases.
The displayed error is the mean of 30 realizations of random stationary graph signals and only one hidden variable is considered.
Also, for each of the three graphs we include the performance of both the joint and the separate estimation.
It can be observed that the recovery of the three graphs improves when a joint approach is followed, showcasing the benefits of exploiting the existing relationship between the different networks.
Furthermore, this experiment confirms that the developed method is also suitable for real applications.

\vspace{-1mm}
\section{Conclusions}\label{conclusions}
\vspace{-1mm}
In this paper, we presented a new method for solving the challenging problem of joint graph topology inference in the presence of hidden nodes.
To tackle this ill-posed inference problem, we assume that (i) the number of hidden nodes $H$ is much smaller than the number of observed nodes $O$; (ii) the observed signals are realizations from a random process stationary in $\bbS^{(k)}$; and (iii) the $K$ graphs are closely related.
Furthermore, we exploit the inherent block structure of the matrices $\bbC^{(k)}$ and $\bbS^{(k)}$ to solve the joint topology inference problem by solving an optimization framework.
A reweighted $\ell_1$ norm to promote sparse solutions is employed, and the stationarity assumption is adapted to the presence of hidden nodes by defining the (unknown) low-rank lifting matrices $\bbP^{(k)}$.
Instead of relying in the nuclear norm, low-rank matrices $\bbP^{(k)}$ are achieved by promoting column-sparsity with the group Lasso penalty.
Moreover, the similarity of the $K$ graphs is leveraged in two ways.
First, for each pair of graphs, we look for matrices $\bbSo^{(k)}$ with a similar edge pattern by minimizing the $\ell_1$ penalty, and second, we look for matrices $\bbP^{(k)}$ with a similar column sparsity pattern.
The proposed method is evaluated using synthetic and real world graphs, and a comparison with other baseline methods based on graph stationarity and on graphical Lasso is provided. 

\newpage

\vfill\pagebreak
\bibliographystyle{IEEE}
\bibliography{citations}

\end{document}